\documentclass[useAMS,usenatbib, referee]{mn2e}
\usepackage{graphicx}
\usepackage{natbib}
\usepackage{times}
\usepackage{float}
\usepackage{epsfig}

\newcommand{\be}{\begin{equation}}
\newcommand{\ee}{\end{equation}}

\title[Dynamical tides in systems containing Hot Jupiters]{
Dynamical tides in exoplanetary systems containing Hot Jupiters: confronting theory and observations}
\author[ S. V. Chernov, P. B. Ivanov, J. C. B. Papaloizou ]{ S. V. Chernov
$^{1}$\thanks{E-mail: chernov@td.lpi.ru (SVCh)} P.B.Ivanov$^{1,2}$\thanks{E-mail:
pbi20@cam.ac.uk (PBI)} and J. C. B. Papaloizou $^{2}$\thanks{E-mail:
J.C.B.Papaloizou@damtp.cam.ac.uk (JCBP)} \\
$^{1}$Astro Space Centre, P.N. Lebedev Physical Institute, 84/32
Profsoyuznaya Street, Moscow, 117997, Russia  \\
$^{2}$ DAMTP, Centre for Mathematical Sciences, University of
Cambridge, Wilberforce Road, Cambridge CB3 0WA }

\begin{document}

\date{Accepted. Received; in original form}

\pagerange{\pageref{firstpage}--\pageref{lastpage}} \pubyear{2010}

\maketitle

\label{firstpage}

\begin{abstract}
We study the effect of dynamical tides associated with the excitation of gravity waves in an interior radiative region of the central star on orbital evolution in observed systems containing Hot Jupiters. We consider WASP-43, Ogle-tr-113, WASP-12, and WASP-18 which contain stars {  on} the main sequence (MS). For these systems there are observational estimates regarding the rate of change of the orbital period. We also investigate Kepler-91 which contains an evolved giant star. We adopt the formalism of Ivanov et al. for calculating the orbital evolution.

For the MS stars we determine expected rates of orbital evolution under different assumptions about the amount of dissipation acting on the tides, estimate the effect of stellar rotation for the two most rapidly rotating stars and compare results with observations. All cases apart from possibly WASP-43 are consistent with a regime in which gravity waves are damped during their propagation over the star. However, at present this is not definitive as observational errors are large. We find that although it is expected to apply to Kepler-91, linear radiative damping cannot explain this dis- sipation regime applying to MS stars. Thus, a nonlinear mechanism may be needed.

Kepler-91 is found to be such that the time scale for evolution of the star is comparable to that for the orbit. This implies that significant orbital circularisation may have occurred through tides acting on the star. Quasi-static tides, stellar winds, hydrodynamic drag and tides acting on the planet have likely played a minor role.
\end{abstract}

\begin{keywords}
hydrodynamics - celestial mechanics - planetary systems:
formation, planet -star interactions, stars: binaries: close,
oscillations
\end{keywords}

\section{Introduction}
In recent years the discovery
of many extrasolar planets orbiting in close proximity to
their central stars has highlighted a situation where tidal interactions
are likely to have been important in determining the formation and
subsequent orbital evolution of the systems
\citep[e.g.][]{Terquem, Barker2009}.

In particular, Hot Jupiters may
be formed by a tidal capture into a highly eccentric orbit followed
by orbital circularisation.
During this process
tidal dissipation both in the planet and the star can be
significant  \citep[see e.g.][and references therein] {Rasio, IP2007, IP2010, IP2011}.
In the latter case the tidal dissipation may lead to the planet merging with the
central star at a later stage of its evolution \citep[e.g.][]{vill}.

An understanding of these processes requires an analysis of the tidal interaction of the
planet with the central  star and its calibration through comparison with observations.
In particular, the  rate of orbital evolution 
can be inferred from observed  orbital period changes
 \citep[see][] {Hebb2009, Hellier2009, HoyerMNRAS2016, Jiang2016, Maciejewski2016}
making such a calibration possible in principle. In order to proceed with this
we consider 
tides raised on the central star by planets on near circular orbits
as these {  are relevant to observed systems}. Orbital circularisation can also occur
but tides raised on the planet could also  play a role there.
We also focus on dynamical tides as we anticipate that quasi-static tides
are unlikely to be important on account of the mismatch of the tidal forcing frequency and the
inverse convective turn over time (see Section \ref{Quasi-static tides} below).
Dynamical tides are found to be associated with the excitation of potentially resonant
gravity or $g$ modes.

\citet{IPCh} (IPCh)   determined  the tidal  response  associated with   the excitation of  a  regular
dense spectrum of normal modes, such as  provided by the low
frequency  rotationally  modified  $g$  modes, by a perturbing tidal potential.  They  obtained
expressions  from which the orbital evolution could be obtained.
These depend on the amount of dissipation present. Two regimes were highlighted.
The  regime of  moderately large  damping 
(MLD)  for which the  excited
waves  are damped before reaching an appropriate
boundary, the centre for a radiative core, and the surface for
a radiative envelope.  In this regime the effect on the orbit  is independent of the
details of the dissipation process.
Note that the same assumption of the validity of MLD regime is implied in the well known theory
of dynamical tidal interactions of \citet{Z1} and \citet{Z2}  which is however, contrary to the IPCh
approach, formally valid only in the asymptotic limit of tidal forcing frequency tending to zero,
and for stars with an idealised structure.
In addition, the expressions for the orbital evolution
are applicable to the case of very weak dissipation where resonant
responses may occur.

It is the purpose of this paper to apply the formalism of IPCh
for determining the effect of dynamical tides on orbital evolution to
observed exoplanet systems containing close orbiting Hot Jupiters.
In particular we consider WASP-43,   Ogle-tr-113, WASP-12,
and  WASP-18 which contain stars {  on} the main sequence (MS)
with Hot Jupiter companions for which there are observational estimates
of, or upper bounds for,  the magnitude of the rate of change of their orbital periods.
In addition we consider the system Kepler-91 which contains a star that
has evolved up the giant branch and a Hot Jupiter companion.

For the MS stars we determine the expected rates of orbital evolution, assuming the orbit to be circular,
either under the assumption that the MLD regime applies,
or under the assumption that damping is weak with the system having evolved so that
the tidal forcing frequency is mid way between neighbouring  potentially resonant normal mode frequencies
and compare results with observations.
We also estimate  the effect of rotation for the two most rapidly rotating cases
by simply allowing for the shift of forcing frequency.
All  cases apart from WASP-43  can be viewed as being  consistent with being in the MLD regime,
however
it has to be emphasised  that in general,  limits on  observational errors are large
so that this is not definitive.
We remark that   although  we find that can be applied  
to the giant in Kepler-91,
 we find that the MLD regime is not  expected to operate in
the MS stars if linear radiative damping is the only available dissipation mechanism.
Thus, a nonlinear mechanism may also  be required.

Kepler-91 is found to be in  a configuration where the time scale for evolution of the star
is comparable to the time scale for evolution of the orbit. In particular this implies that significant
orbital circularisation may have occurred   as a result of tides in the past.

The plan of this paper is as follows. We begin by giving  some basic equations and definitions in
Section \ref{Basiceq}, moving on to present the equations governing
the  evolution of orbital parameters induced by tides in Section \ref{Tidev}.
We evaluate the decay rate due to radiative damping of the high order $g$ modes,  that are expected to be excited in
tidal interactions,
in Section \ref{Modedecay}, developing  a criterion
for the excited modes  to be in the  the  moderately large damping  (MLD) limit
in Section \ref{MLDcrit}.

 We describe the procedure we use to obtain
solutions of the equations determining orbital  evolution  under tides  in  Section \ref{Eqsol},
giving
properties of the  stars, and the orbits of their planetary companions,  for the systems
we study in  this paper in Sections \ref{Starprops},
\ref{MSstars},
and \ref{stellarotation}.
The   decay rates of the  normal modes expected to be excited by tides  acting on  the stars {  on} the main sequence,
and the  appropriateness  of the MLD limit for them,  are  then considered in Section
\ref{NMdecayratesMS}.

We move on to describe our modelling of the evolved  star in
Kepler-91 in Section \ref{Kepler91stellarmodel}. We consider the
properties of appropriate  normal mode decay rates in  Section \ref{NMdecayratesKepler91},
establishing that the MLD limit applies to the current configuration of star and planet.

We determine the expected  tidal evolution for the systems we study  in
Section \ref{Tidalevolution}, giving results for systems with stars {  on}  the main sequence
in Section \ref{MSresults} and for
the orbital evolution of Kepler-91b  in Section \ref{SKepler91b}.
For Kepler-91 we consider effects due to quasi-static tides, a possible stellar wind, and hydrodynamic drag, which are discussed in Sections
\ref{Quasi-static tides},
\ref{Stellar wind}, and
\ref{hydrodynamic drag}, respectively.
Finally, we discuss our results  and conclude in Section
\ref{Discuss}.

\section{Basic definitions and equations}\label{Basiceq}

We consider a binary system consisting of a star of
mass $M$ with radius $R_*$ that is  orbited by a planet of mass $m,$ the mass ratio being  $q=m/M.$
The star may in principal be rotating, but we  assume  that angular velocity of  rotation is much
smaller than the characteristic  angular frequency of the normal modes   excited by tides,
the latter being expected to be comparable to the orbital mean motion.

The planet moves around 
the star on an approximately circular orbit with  period $P_{orb}$  of  the order of  days.
 The orbital
semi-major axis is $a = (GM/\Omega_{orb}^2)^{1/3}$, where $G$ is gravitational
constant and $\Omega_{orb}={2\pi / P_{orb}}$ is the  Keplerian mean motion or  angular velocity. The orbital eccentricity, $e$,
is such that $e\ll 1$.
We assume that the  stellar rotation axis is aligned with the direction of orbital angular momentum, and that the star rotates uniformly and relatively slowly with angular velocity, $\Omega,$ such that $\Omega \ll \Omega_{orb} \ll \Omega_*,$ where we have introduced  a characteristic  frequency associated with the star,
$\Omega_*=\sqrt{GM/ R_*^3}$.

\subsection{Evolution of orbital parameters induced by tides}\label{Tidev}
Following \citet{IPCh} (IPCh) we write the equations governing the evolution of the semi-major axis $a,$ and the eccentricity, $e$,  as
\begin{equation}
{\dot a\over a}=-{2\over T_{a}} \quad {\rm and }\quad {\dot e\over e}=-{1\over
T_{e}}, \label{t1}
\end{equation}
where $T_a$ and $T_e$ are characteristic timescales for  the evolution of the semi-major axis and
eccentricity, respectively.
For a non-rotating primary and  near circular orbit  IPCh give  expressions
for $T_a,$ and $T_e$ in terms
of quantities characterising the orbit and the star
in the form
\begin{equation}
T_{a}={40 T_* \over 3\pi}\left [{ Q}^{-2}D^{-1}\right ]_2 \quad {\rm and}\quad T_e={20\over \pi}T_*/{\cal{ F}}(\Omega_{orb}).
\label{t11}
\end{equation}
Here
\begin{equation}
{\cal{ F}} (\Omega_{orb})= \left \lbrace {49\over 18}\left
[{ Q}^2D\right ]_{3}-{3\over 4}\left [{ Q}^2D\right
]_{2}+\frac{15}{2}\left [{ Q}^2D\right ]_{1}\right \rbrace,
 \label{t17}
\end{equation}
and
\begin{equation}
T_*={1\over 16\pi^3}{\left(1+q \right)^{5/3}\over q
}\left({P_{orb}\Omega_*\over 2\pi}\right)^{4/3}\left|{d\omega_j\over dj}\right |_{j=j(k)} P_{orb}^2.
\label{t9}
\end{equation}

Quantities  enclosed in square
brackets $[..]_{k},$ with the subscript, $k,$ being an integer,
are functions of  an eigenfrequency $\omega_{j=j(k)}.$
This is found by evaluating the frequency offset $\Delta \omega_j$
\begin{equation}\Delta \omega_j= k\Omega_{orb}-2\Omega -\omega_j,
\label{t9n}
\end{equation}
for each normal  mode eigenfrequency,   $\omega_j$ $(j = 1, 2, ..)$
and choosing  $j(k)$  to be the value of $j$ for which the magnitude of the  frequency offset is minimal.
This corresponds to selecting  the particular mode  that is closest to being resonant with a component of
the perturbing tidal potential.
Note that only the modes that are actually excited for a specified $k$ should be considered in this determination.
Let us stress that in practice  we always have $|\Delta \omega_{j=(k)}| \ll k\Omega_{orb}-2\Omega \approx \omega_{j=j(k)}$,  corresponding  to a sufficiently
dense spectrum of eigenmodes.

The quantity $Q_{k}$ in (\ref{t11}) and (\ref{t17}) is
the overlap integral evaluated for the normal mode with $\omega = \omega_{j=j(k)}$
(see equation (47)  of IPCh and the  discussion that follows there ).
In principle,  stellar rotation affects the form of expressions
(\ref{t11}-\ref{t9}), see IPCh. However, as we have indicated above,  we  consider only the case of a relatively slowly rotating star, and, therefore, take into account only the dominant effect for high order modes, namely, the frequency
shift due to the presence of $\Omega $ in equation (\ref{t9n}) \citep[e.g.][]{Goodman} .

 We set $\left| d\omega_j/ dj \right|_{j=j(k)}  \equiv \omega_{j+1}-\omega_j,$ being the frequency difference between two
neighbouring modes which are  such that $k\Omega_{orb}-2\Omega $ lies in between $\omega_j $ and $\omega_{j+1}$.
 This  is written as a derivative which is appropriate in the limit of modes of high  order  $(j  \gg 1).$
It was explicitly evaluated
in IPCh for the case of high order $g$-modes in Sun-like stars. In this case
the expression (\ref{t9}) can be rewritten as
\begin{equation}
T_*={\left(1+q \right)^{5/3}\over \sqrt{6} q
}\left({P_{orb}\Omega_*\over 2\pi}\right)^{4/3}\left( \int_{\cal D}{dr\over r}N\right)^{-1},
\label{tn9}
\end{equation}
where $N$ is  the Brunt - V$\ddot {\rm a}$is$\ddot
{\rm a}$l$\ddot {\rm a}$ frequency and the integral is over a domain ${\cal D}$  which defines a radiative region in which  $g$ modes can propagate.
Note that we assume that $k=2$ in \ref{tn9} and below.
Thus, we use (\ref{tn9}) when considering stars with radiative interior
and convective envelope, and in all other cases a more general expression
(\ref{t9}) is used.

We remark that equations (\ref{t11}) - (\ref{t9})  with $D=1$  corresponding to the  limit of 'moderately large viscosity',      {  or moderately  large dissipation}  
 (MLD) described below, can be found from
equations (128),  (131), (137) and (138) of  IPCh.
The function $D$ accounts for the influence of mode damping rate, $\gamma$, assumed to originate
from the action of either {  linear radiative damping} \footnote{ But note that any dissipative process that results in a radiation boundary condition for waves propagating through
the  radiative domain of interest results in behaviour
corresponding to  the MLD regime  (see IPCh).}  or non-linear effects.  Note that  $\gamma$ replaces $\nu_{j(k)},$ the decay rate of a normal mode,  as used in IPCh.
Explicitly, $D$ has the form
\begin{equation}
D={\sinh (\pi \kappa )\cosh (\pi \kappa )\over \sinh^2 (\pi \kappa) + \sin^2 (\pi \delta)},
\label{en1a}
\end{equation}
where $\delta = |\Delta \omega_j | /| d\omega_j/dj|_{j=j(k)}$,
$\kappa = \gamma / |d\omega_j/dj|_{j(k)}$.
We remark that $D$  may be written as
 $D=(\kappa / \pi ){\cal A}_{\kappa}$, where ${\cal A}_{\kappa}$ is given by equation (44) of  IPCh.

When $\kappa > 1,$
$D\approx 1$. In this MLD  limit, tidal evolution does not
depend on the mode damping rate. Physically, this corresponds to a situation when a wave
packet excited in a star by tides decays in course of its propagation over the star. This limit was implied in the old theory of dynamic tides \citep{Z1,Z2}.
Expressed quantitatively, the condition to be in this regime is that the time  for a gravity wave to propagate through the radiative region should exceed the mode damping
time ( for more detail see below). When this is not satisfied, the full expression for $D$ must be used in (\ref{t11})  and (\ref{t17}).
In what follows we discuss   whether
the MLD limit applies,
both from the theoretical point of view,  and whether this is supported by observational data on the orbital evolution of exoplanetary systems containing
Hot Jupiters, under the assumption that this evolution  is caused by tides.

We first make  theoretical
estimates in order to determine the applicability  of  the MLD limit
to systems with exoplanets.
We find that  although the  mechanism  of  radiative damping  allow us to justify it  in the case of  evolved stars
this is inadequate for our models of main sequence stars.
In the latter case some non-linear   mechanism of  mode energy dissipation
has to  be invoked to justify it.
In the absence  of such a mechanism the opposite limit $\kappa \ll 1$ corresponding to weak dissipation
is valid. Then we have
\be
D\approx {\pi \kappa \over (\pi \kappa)^2 +  \sin^2 (\pi \delta)}\label{smallkap}.
\ee
We see from the tidal evolution equations (\ref{t1})-(\ref{t9})  that in this case tidal evolution rate is  proportional to the mode damping rate
unless the system is very close to an exact resonance such that $\delta \ll \kappa. $
This can be extremely rapid, but only near the centre of a resonance. In such cases
{  unless resonances can be maintained by a locking process
 \citep[see][] {Witte1999,Witte2002,Fuller2016}}  systems would rapidly
evolve away from such a configuration so that they would be most likely to be found between resonances.
From the definitions just below equation  (\ref{en1a}), we see that mid way between resonances $\delta=1/2$ and accordingly in the limit of weak dissipation
\be D= \pi\kappa = \pi\gamma / |d\omega_j/dj|_{j = j(k)}.\label{WEAKR}\ee
This is the factor by which the evolution rates assuming that the   MLD  limit holds
has to be multiplied when the system is in fact in the regime of very weak  dissipation.

{  If resonances can be maintained through locking, tidal dissipation and evolution can be very rapid. This cannot occur in the MLD regime as standing waves 
and strong resonances
cannot be set up. We remark that  \citet{Witte1999,Witte2002}
find that this mechanism is effective mainly for eccentric orbits and  then it
 can lead to efficient
circularization. For short period planets in near circular orbits,
they find that orbital  period evolution occurs on 
long  time scales and is very much slower than
 expected were the MLD regime to  operate.} 
We now go on to develop estimates  for the magnitude of 
radiative damping which may be used to determine  the regime of dissipation
that applies and evaluate $D$ when dissipative effects  are very  small.

\subsection{Decay rate due to radiative damping}\label{Modedecay}
  The decay rate is given  of a $g$ mode with frequency $\omega$ is given by \citet{Unno89}  as
\be
\gamma = \frac{1}{2\omega^2}\frac{\int_V (\delta\rho^*/\rho)(\Gamma_3-1)\nabla\cdot{\bf F}' d\tau }
{\int_V \rho |{\mbox{\boldmath{$\xi$}}}|^2d\tau}.\label{decay}
\ee
 The density is $\rho$ and the radiation flux is {\bf F}.  Eulerian perturbations are donated with a prime and the Lagrangian variation is indicated by a preceding $\delta$.
The normal modes we consider are such that the angular dependence of $\rho'$ is through a spherical harmonic with indices  $l=|m|=2.$
This will be taken as read in what follows.
All  quantities in the integrals may in principle be expressed in terms of $\mbox{\boldmath{$\xi$}}.$ This  is facilitated in the quasi-adiabatic
 approximation. We then have
\be
\delta\rho =\rho' + \mbox{\boldmath{$\xi$}}\cdot\nabla\rho = {1\over \Gamma_1 P}\rho( P' + \mbox{\boldmath{$\xi$}}\cdot\nabla P),
\ee
where $P$ is the pressure and $\Gamma_1$ the first adiabatic exponent.
We  remark that for  g modes in the asymptotic  low frequency limit we can set $P' =0$ in the above,  then after use of hydrostatic equilibrium we obtain
\be
\delta\rho
= {1\over \Gamma_1 P}\rho \mbox{\boldmath{$\xi$}}\cdot\nabla P= - \rho^2 (\Gamma_1P)^{-1}\mbox{\boldmath{$\xi$}}\cdot{\bf g},
\ee
where ${\bf g}$ is the acceleration due to gravity.
We also have
\be
\delta T =T' + \mbox{\boldmath{$\xi$}}\cdot\nabla T
= ( \Gamma_2 -1)T( \Gamma_2P)^{-1}( P' + \mbox{\boldmath{$\xi$}}\cdot\nabla P),
\ee
where $T$ is the temperature and $\Gamma_2$ and $\Gamma_3$ are the second and third adiabatic exponents respectively.
In the low frequency asymptotic limit
this similarly yields
\be
\delta T
= -( \Gamma_2 -1)\rho T( \Gamma_2P)^{-1} \mbox{\boldmath{$\xi$}}\cdot{\bf g}.
\ee
The radiative flux is given by
\be
{\bf F} =- \frac{4acT^3}{3\kappa \rho}\nabla T,
\ee
where $\kappa$ is the opacity, $a$ is the radiation
constant and $c$ is the speed of light.
Linearizing and noting that
as very short radial wavelengths are  expected
in the limit of low frequency g modes,
we may  retain only the  highest order radial  derivatives of perturbations,  we may write
\be
{\bf F}' \rightarrow - \frac{4acT^3}{3\kappa \rho}\nabla T'
\ee
and
\be
\nabla\cdot {\bf F}' \rightarrow - \frac{4acT^3}{3\kappa \rho}\nabla^2 T' \rightarrow  \frac{4acT^3}{3\kappa \rho}k_r^2 T',
\ee
where $k_r$ is the radial wavenumber.
Making use of the above approximations in (\ref{decay})  we estimate the damping rate through
\be
\gamma \omega^2 \int_V \rho |{\mbox{\boldmath{$\xi$}}}|^2d\tau=
2\int_V \frac{\rho\mbox{\boldmath{$\xi$}}^*\cdot{\bf g}(\Gamma_3-1)ack_r^2T^4
\mbox{\boldmath{$\xi$}}\cdot{\bf g}( - \nabla
+\nabla_{ad})
} {3 \Gamma_1P^2\kappa }d\tau ,\label{decay1}
\ee
where
$\nabla = d\log T/d\log P$ and $\nabla_{ad}= (\Gamma_2-1)/\Gamma_2.$
Note that the {   integral on the right hand side} contains only the radial component of the displacement, $\xi_r,$ whereas the {  integral on the left hand side contains in addition}
$\xi_{\perp}~\equiv~|{\mbox{\boldmath{$\xi$} }}- \xi_r{\hat{\bf r}}|\equiv |{\mbox{\boldmath{$\xi$} }}_{\perp}|$
and we expect $\xi_{\perp} \gg |\xi_r|.$
For g modes in the asymptotic low frequency  limit we have $|k_r\xi_r|^2 \sim   |\nabla\cdot {\mbox{\boldmath{$\xi$} }}_{\perp}  |^2$
 and  we may set $|{\mbox{\boldmath{$\xi$}}}|^2 \sim( k_r^2 r^2/(l(l+1)) +1 )|\xi_r|^2.$
\noindent Using this in (\ref{decay1}), we get
\be
\gamma \omega^2 \int_V \rho (r^2 k_r^2+l(l+1)) |\xi_r|^2d\tau=
2l(l+1)\int_V \frac{\rho|\xi_r|^2 g^2(\Gamma_3-1)ack_r^2T^4
( - \nabla
+\nabla_{ad})
} {3 \Gamma_1P^2\kappa }d\tau .\label{decay2}
\ee

To proceed further, we note that from the WKBJ approximation (see eg. IPCh)
\be
k_r^2= \frac{l(l+1)}{r^2} \left( \frac{N^2}{\omega^2}-1\right)\label{kr}
\ee
and that for real $k_r,$
\be
\xi_r \propto \rho^{-1/2}r^{-3/2}(N^2/\omega^2-1)^{-1/4}\exp({\rm i}\Phi).\label{xir}
\ee
Here the proportionality factor includes the angular dependence of the mode
through a spherical harmonic
and the phase
\be
\Phi = \int^{r}_{r_0}k_rdr+ C_0,
\ee
with  $r_0$ and $C_0$ being constants depending on the boundary conditions.

Using (\ref{kr}) and (\ref{xir}) in (\ref{decay2}) we obtain an expression from which the
decay rate may be readily calculated in the form
\be
\gamma \int_{\cal D} \frac{ N^2 }{r(N^2/\omega^2-1)^{1/2}}dr=
2l(l+1)\int_{\cal D} \frac{ g^2(\Gamma_3-1)acT^4
( - \nabla
+\nabla_{ad})(N^2/\omega^2-1)^{1/2}
} {3 \Gamma_1 r^3 P^2\kappa }dr .\label{decay3}
\ee
In this work the domain of integration  ${\cal D}$ is restricted to the wave  propagation  region
in the interior radiative region for which $k_r^2 > 0$ and we recall that for the modes of interest $l=2.$
\subsection{Criterion for being in the MLD limit}
\label{MLDcrit}
In order for the quasi-adiabatic approximation to be applicable to a mode,
we require $\gamma/\omega \ll 1.$
However, the condition
for a disturbance excited externally to be damped before it passes 
through the radiative region where $g$ modes can propagate is that the
time to move across
the region with the group velocity be $\gg \gamma^{-1}.$
This condition for being in the  regime of MLD
 can be satisfied while  the quasi-adiabatic approximation is valid.
 An expression for  estimating it can be found by first noting that
 in the WKBJ approximation the normal modes satisfy (see eg. IPCh)
 \be
 \int_{\cal D} k_r dr = n\pi + \delta,\label{WKBJM}
 \ee
  where $n$ is a positive  integer and $\delta$ is a constant determined through the boundary conditions
  and WKBJ connection formulae.  Making use of (\ref{kr}), from (\ref{WKBJM})  we find that
  \be
\frac{ d\omega}{dn} \int_{\cal D}\frac{\partial  k_r}{\partial \omega} dr = \pi .\label{WKBJVG}
 \ee
Thus
 \be
\pi \gamma /(d\omega/{dn})  = \gamma \int_{\cal D}\left(\frac{\partial  \omega}{\partial k_r}\right)^{-1} dr .\label{WKBJVG1}
 \ee
The right hand side of (\ref{WKBJVG1}) is the product of the
time to propagate through the region with the group velocity and the mode decay rate.
Accordingly we shall adopt the criterion to be in the moderately dissipative regime
that  $\gamma /(d\omega/{dn}) > 1.$ When it is marginally satisfied a wave pulse has an amplitude
reduction by a factor, $\exp(\pi),$  on propagating through the propagation  zone.
We remark that 
when $\gamma /(d\omega/{dn}) = 1$, at the centre of a resonance, from (\ref{en1a}) we find
consistently that   $D= \coth\pi =1.004. $

\section{Solution of the equations determining the evolution of the orbit under tides }\label{Eqsol}
Equations (\ref{t1})-(\ref{t9}), which govern the evolution  of the orbital elements,
 are solved numerically according to the following procedure \citep{Chernov}.
We  initially generate a stellar model with parameters appropriate to a particular system we wish
to study. We  then calculate eigenfrequencies and overlap integrals
corresponding to normal modes with frequencies in the range required to evaluate the terms in equations
(\ref{t1})-(\ref{t9}), for the range of orbital periods of interest,
  using the approach described
in  \citet{CD1998}.
Since these are relatively low frequency modes they belong to $g$-mode branch of stellar pulsations.
For these modes,  self-gravity plays a minor role and we neglect it, thus adopting  the Cowling  approximation \citep{Cowling}.

We then integrate  equations (\ref{t1}-\ref{t9}) numerically in order to determine the tidal evolution of the orbit
after having specified initial values for the orbital  period, eccentricity and age of the system.
 In this paper we consider tidal evolution in systems containing both main sequence
stars and  evolved stars that have moved  off the main sequence and along   the giant branch.
In the latter
situation  the stellar  evolution time scale   can  be  comparable with the  time scale for orbital evolution
under  tides.
 For such cases we generate a grid of models with different ages so that
 time derivatives of $a$ and $e$  can be calculated  for a stellar  model
 that self-consistently has the correct age.

\section{Properties of the  stars and their planetary companions in the systems studied}\label{Starprops}

\subsection{Main sequence stars}\label{MSstars}

In this section we consider the three  systems containing main sequence stars,  WASP-43,  Ogle-tr-113 and WASP-12 in some detail.
Each of these  systems contains a planet
with   mass  of  around one Jupiter  mass. There is also a measured rate of change of orbital period with time.
In order to construct  models we use the  publicly available stellar evolution code MESA (Modules for Experiments in Stellar Astrophysics),
\citep[see][ and  http://mesa.sourceforge.net/.]{Paxton11,Paxton13,Paxton15}
We give the  main observational  parameters for the  stars we  have considered  in table \ref{wasp43}
together with the corresponding quantities for  our associated  numerical models. In table
\ref{wasp43b}
 we show masses, radii, orbital periods, and their published  observed  rates of change,
$\dot P_{orb}$, together with corresponding error bars.

Additionally, we consider the system WASP-18. For this system only an upper limit for orbital change is available, see \cite{wilkins}. We check whether or not this upper limit is consistent with the assumption that MLD
regime operates. Main observational parameters of the star and
the planet as well as properties of our two numerical stellar models are
shown below in tables \ref{wasp43} and
\ref{wasp43b}. In order to obtain the   upper limit for  $\dot P_{orb}$ given in  table  \ref{wasp43b}, we
use   the   estimate  of \cite{wilkins} that the effective  modified tidal quality factor, $Q^{'}$, should be larger than $10^6$ together with the expression
from for the rate of change of semi-major axis due to tides  given by  \cite{birk}.
This gives $|\dot P_{orb}| < 0.02s/yr$.

\begin{table}
\begin{minipage}{\textwidth}
\begin{tabular}{|c|c|c|c|c|c|}
\hline
   & $M$ & $R_*$ & [Fe/H] & $T_{\rm eff}$ & age\\
\hline
 WASP-43 & 0.717$\pm$0.025$M_\odot$ & 0.667$\pm$0.01$R_\odot$ &-0.01$\pm$0.012 & 4520$\pm$120 K&  $> 0.4Gyr$ \\
\hline
  Model & 0.717$M_\odot$ & 0.667$R_\odot$ & -0.011 & 4384 K & 0.75Gyr\\
\hline
 Ogle-tr-113 & 0.78$\pm$0.02$M_\odot$ & 0.765$\pm$0.025$R_\odot$ & 0.14$\pm$0.14 & 4751$\pm$130 K & 0.7 \\
\hline
 Model  & 0.78$M_\odot$ & 0.721$R_\odot$ & 0.14 & 4753 K & 0.69Gyr\\
\hline
 WASP-12 & 1.35$\pm$0.14$M_\odot$ & 1.599$\pm$0.071$R_\odot$ & 0.3$\pm$0.1 & 6300$\pm$150 K&  $1.7\pm 0.8Gyr$ \\
\hline
  Model A& 1.32$M_\odot$ & 1.631$R_\odot$ & 0.243 & 6445 K & 1.53Gyr\\
\hline
 Model  B& 1.32$M_\odot$ & 1.696$R_\odot$ & 0.243 & 6350 K & 1.77Gyr\\
\hline
 WASP-18 & 1.25$\pm$0.13$M_\odot$ & 1.216$\pm$0.067$R_\odot$ & 0.0$\pm$0.09 & 6400$\pm$100 K&  $1.0\pm 0.5Gyr$ \\
\hline
  Model A & 1.24$M_\odot$ & 1.245$R_\odot$ & 0.0 & 6279 K & 0.68Gyr\\
\hline
  Model B & 1.24$M_\odot$ & 1.358$R_\odot$ & 0.19 & 6398 K & 1.08Gyr\\
\hline
\end{tabular}
\caption{Upper line:
Mass, radius, metallicity, effective temperature and an estimate of the age of the  stars,  \newline
WASP-43  \citep[see][]{HoyerMNRAS2016,Jiang2016},  \newline
  Ogle-tr-113 \citep[see][]{Adams2010, HoyerMNRAS2016},\newline
  WASP-12 \citep[see eg.][]{Hebb2009,Maciejewski2016} and
 WASP-18 \citep[see ][]{Hellier2009}.\newline
   In each case the same quantities obtained  from our numerical  models of the stars
are given below \newline
the observational parameters. Note that for WASP-12 and WASP-18 there
are two models A and B.}
\label{wasp43}
\end{minipage}
\end{table}
\begin{table}
\begin{minipage}{\textwidth}
\begin{tabular}{|c|c|c|c|c|c|}
\hline
 & $m$ & $R_{\rm pl}$ & $P_{orb}$ & $dP_{orb}/dt$ \\
\hline
  WASP-43b &2.052$\pm$0.0534$M_J$ & 1.036$\pm$0.019$R_J$ & 0.8135 d & -0.0289$\pm$0.0077 s/yr; -0.00002$\pm0.0066$ s/yr\\
\hline
 Ogle-tr-113b & 1.24$\pm$0.17$M_J$ & 1.11$\pm$0.05$R_J$ & 1.4325 d & -0.001$\pm$0.006 s/yr\\
\hline
 WASP-12b & 1.404$\pm$0.099$M_J$ & 1.736$\pm$0.092$R_J$ & 1.0914 d & -0.0256$\pm$0.0040 s/yr; -0.029$\pm$ 0.003s/yr\\
\hline
 WASP-18b & 10.30$\pm$0.69$M_J$ & 1.106$\pm$0.072$R_J$ & 0.9415 d & $< 0.02s/yr$  \\
\hline
\end{tabular}
\caption{Mass, radius, orbital period and the observed rate of change of orbital period of
 exoplanets,  
 WASP-43b 
\citep[see][]{HoyerMNRAS2016,Jiang2016}, Ogle-tr-113b 
\citep[see][]{HoyerMNRAS2016}, WASP-12b 
 \citep[see][]{Hebb2009,Maciejewski2016,patra} and  WASP-18b 
\citep[see][]{Hellier2009}.
In the case of  WASP-43b
the  larger value of the orbital decay rate quoted was  
 taken from \citet{Jiang2016} and
 the smaller value was obtained   from \citet{HoyerAJ}.}
\label{wasp43b}
\end{minipage}
\end{table}

\begin{figure}
\begin{center}
\vspace{8cm}\includegraphics{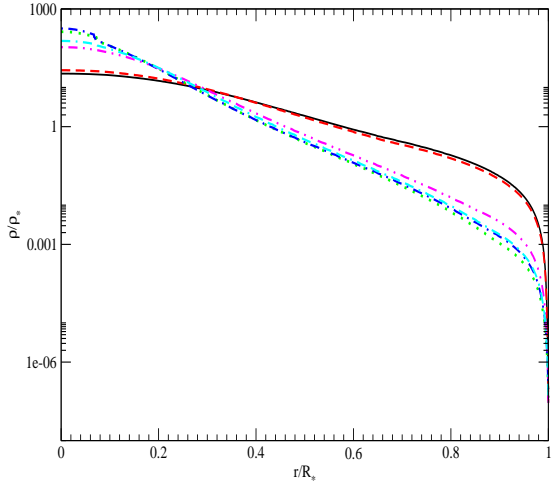}
\end{center}
\vspace{0.5cm} \caption{Distributions of the ratio of the  density $\rho$ to the  mean density $\rho_*=3M/ (4\pi R_*^3)$ shown as
functions of the  dimensionless radius $r/R_*$. Solid, dashed, dotted, dot dashed, dot dot dashed and dot dashed dashed lines correspond to WASP-43,  Ogle-tr-113b, models A and B of WASP-12 and  models A and B of WASP-18, respectively. } \label{density}
\end{figure}




\begin{figure}
\begin{center}
\vspace{8cm}\includegraphics{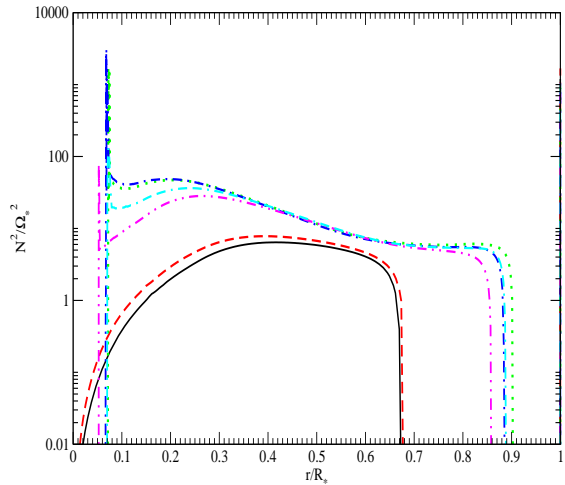}
\end{center}
\vspace{0.5cm} \caption{Same as for Fig. \ref{density}, but the distributions of the  square of the Brunt - V$\ddot {\rm a}$is$\ddot
{\rm a}$l$\ddot {\rm a}$ frequency are shown.} \label{N2}
\end{figure}


We show the dependence of the density and  Brunt - V$\ddot {\rm a}$is$\ddot
{\rm a}$l$\ddot {\rm a}$ frequency
on radius  for each of the stellar models in Figs. \ref{density} and \ref{N2}.
One can see that the models of WASP-12 and WASP-18 are
more centrally condensed than those of the others.
The difference between the models of  WASP-12 and WASP-18 and those of WASP-43
and  Ogle-tr-113b is even more prominent when the respective distributions of the  Brunt - V$\ddot {\rm a}$is$\ddot
{\rm a}$l$\ddot {\rm a}$ frequency are compared.
 The  Brunt - V$\ddot {\rm a}$is$\ddot
{\rm a}$l$\ddot {\rm a}$ frequency is expressed in units of  the inverse of the characteristic  stellar dynamical time scale $\Omega_*=\sqrt{GM/ R_*^3}$.
While the latter models have convective envelopes and radiative cores, and are in general similar to solar models the former models have small convective cores and radiative envelopes. This distinction
is a consequence of the difference in stellar masses and is expected for stars {  on}  the main sequence.
The masses of  WASP-43 and Ogle-tr-113b are approximately equal to
$0.7M_{\odot}$, while WASP-18 and WASP-12 are significantly more massive, having masses  $M\approx 1.2-1.3M_{\odot}$.

\subsubsection{Stellar rotation}\label{stellarotation}
 For Ogle-tr-113, WASP-12 and WASP-18 we use the data on projected rotational velocities listed in \cite{Sch}  and assume that the inclination angle
between rotational axes and the line of sight is close to $\pi/2$. We  thus  respectively obtain rotational periods, $P_r$, approximately equal to
7.79d, 37d and 5.39d for these systems. In case of WASP-43
we use the result quoted in \cite{hell11} that $P_r\approx 15.6d$. Of these
the  rotation periods of Ogle-tr-113 and WASP-18 are closest to
the orbital period, thus indicating that  for these two systems
 is there a possibility that  the effect of rotation  could weaken tidal interactions  appreciably.
  Accordingly, we consider this effect only for these two systems  below.

{  In addition we remark that there are indications that 
the orbit of WASP-12 is strongly misaligned
with the stellar equatorial plane \citep{Albrecht2012}. This means 
that additional tidal effects
 to those
we consider can  play a role. However, the stellar rotation period is estimated to 
exceed the orbital period by  more than an order of magnitude. Accordingly, as indicated above, 
it should be reasonable to neglect rotation when considering orbital decay.
Nonetheless the misalignement will result in tidal forcing
 associated with azimuthal mode number, $m=1,$ in addition to that
with $m=2$ considered in this paper.  This  will excite stellar modes
 leading to  dissipation
that is expected to cause evolution towards alignment \citep[see eg.][] {Papaloizou1982}.
As long as tides remain linear this effect should be decoupled from orbital decay.} 
\begin{figure}
\begin{center}
\vspace{16cm}\includegraphics{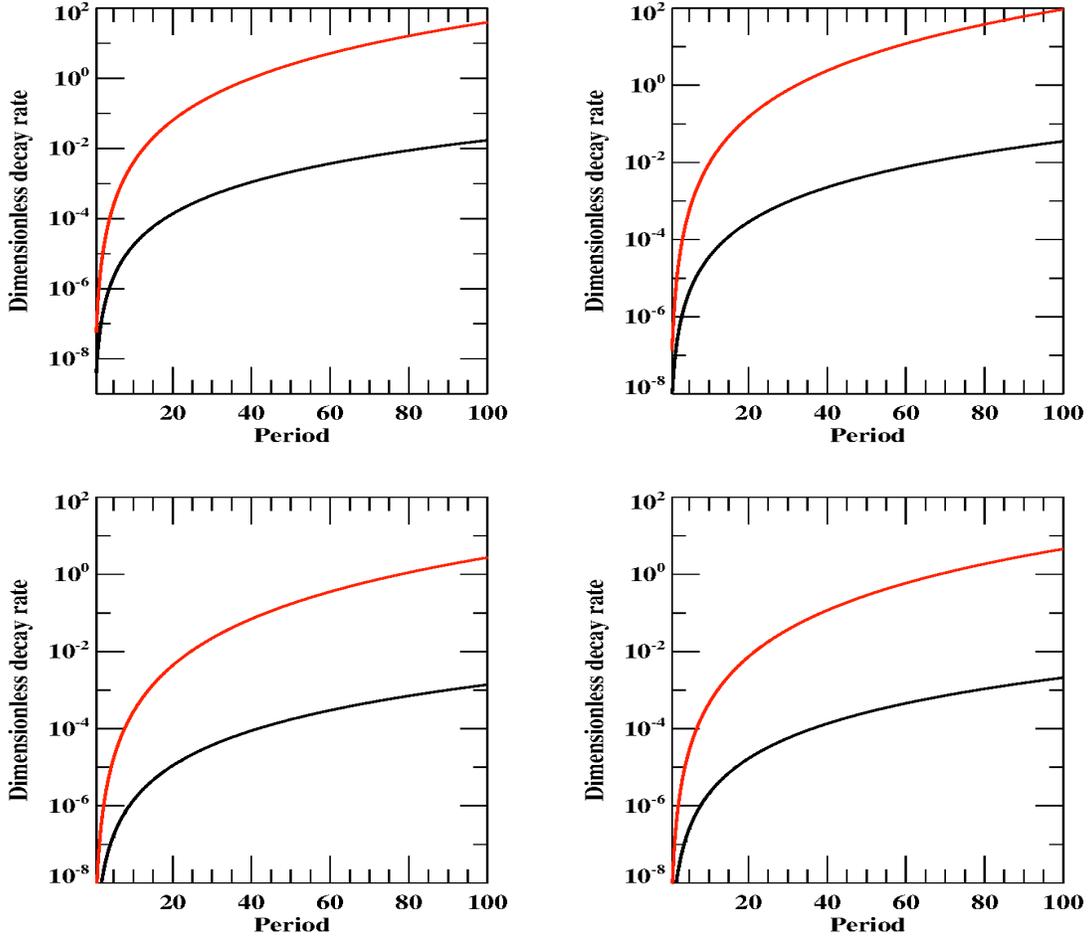}
\end{center}
\vspace{0.5cm} \caption{   Each of the panels shows the ratio of the decay rate of a normal mode to its angular frequency,
 $\gamma/\omega$  ( black curve ) and the ratio of the decay rate of a normal mode to the mode angular frequency interval,
$\gamma/(d\omega/dn)$ (red curve) for stellar models in the vicinity of the main sequence, 
as functions of the orbital period $4\pi/\omega$ in days.
The  upper left panel corresponds to  WASP-18, the
upper right panel to WASP-12, the lower left panel to WASP-43  and the lower right panel to
 Ogle-tr-113  respectively.} \label{No}
\end{figure}
\subsubsection{Properties of normal mode decay rates}\label{NMdecayratesMS}
Following the procedure outlined in sections \ref{Modedecay} and \ref{MLDcrit} we evaluated the normal mode decay rates as a function
of forcing frequency, $\omega,$ corresponding to the excited mode frequency.  This is twice the orbital angular velocity in the case of a circular orbit, which will be assumed
for the purpose of estimating whether the MLD limit applies in this section.

The ratio of the decay rate of a normal mode to its angular frequency, $\gamma/\omega,$ and
 the ratio of the decay rate of a normal mode to the mode angular frequency interval, $\gamma/(d\omega/dn)$
is shown for  main sequence models as a function of the  putative orbital period $4\pi/\omega$ in Fig. \ref{No}.
Note that these are considered as continuous functions of $\omega$, 
even though the normal modes take on discrete values.
However, the frequency interval separating consecutive modes is small such that viewing
the relevant quantities as continuous functions is reasonable.

The models considered are for  WASP-43, Ogle-tr-113, model A for WASP-12
and  model A  for WASP-18. We remark that models  B  give very similar results to models  A.
We see that the  models for WASP-43  and  Ogle-tr-113 produce similar results as do the models for WASP-12
and WASP-18. The latter pair have values for  $\gamma/\omega,$ and  $\gamma/(d\omega/dn)$ that are characteristically $30$ larger at a given orbital
period than those appropriate to the former pair.   For all models  and periods less than $100$ days, $\gamma/\omega < \sim 0.03,$ indicating validity of the
quasi-adiabatic approximation. However note that this quantity $<\sim 10^{-8}$ at periods of, $\sim 3 d,$ characteristic of Hot Jupiters.
 The quantity  $\gamma/(d\omega/dn)$ that we use to indicate  whether the MLD regime applies exceeds unity only for periods exceeding about $80 $  days
in the case of WASP-43  and  Ogle-tr-113  and for periods exceeding  about $35$ days in the case of   WASP-12
and WASP-18.  Thus the MLD limit does not apply to any of these systems for the period range appropriate to Hot Jupiters if linear radiative dissipation alone is considered.

\subsection{Kepler-91: an example of an evolved star}\label{Kepler91stellarmodel}

\begin{figure}
\begin{center}
\vspace{8cm}\includegraphics{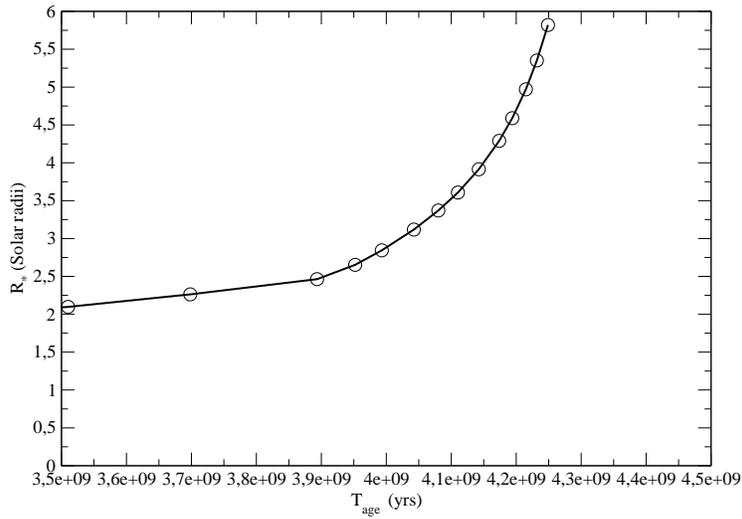}
\end{center}
\vspace{0.5cm} \caption{The evolution of the  radius of Kepler-91 (shown in units of the Solar radius) as a function of time.
 Circles show the positions of particular stellar models used in our computations.} \label{RT}
\end{figure}

We now consider Kepler-91 which is an example of a star which  has evolved off main sequence to move
along  the giant branch. The main parameters of the observed  star  are summarised in table
\ref{Kepler91}
 and the physical and orbital parameters of its companion close-in planet,
Kepler-91b, are given in table \ref{Kepler91b}
The evolution of the  radius of Kepler-91 as a function of time is illustrated in
Fig. \ref{RT}.    It will be seen that  Kepler-91 is currently evolving with a rapidly increasing radius.
 Therefore, as indicated above it is important to consider a set of models with
different ages, $T_{age}$, and to  calculate the overlap integrals and orbital evolution rates
corresponding to  this set of models.
\begin{table}
\begin{tabular}{|c|c|c|c|c|c|}
\hline
   & $M$ & $R_{*}$ & [Fe/H] & $T_{\rm eff}$ & age\\
\hline
 Kepler-91 & 1.31$\pm$0.1$M_\odot$ & 6.3$\pm$0.16$R_\odot$ & 0.11$\pm$0.07 & 4550$\pm$75 K & 4.86$\pm$2.13Gyr \\
\hline
 Model  & 1.31$M_\odot$ & 6.30$R_\odot$ & 0.10 & 4735 K & 4.26Gyr\\
\hline
 HD32518 & 1.13 $\pm$ 0.18 $M_\odot$ & 10.22 $\pm$ 0.87$R_\odot$ & -0.15 $\pm$ 0.04 & 4580 $\pm$ 70 K & 5.83 $\pm $2.58 Gyr \\
\hline
  Model & 1.13$M_\odot$ & 10.20$R_\odot$ & -0.15 & 4612 K & 6.76Gyr\\
\hline
\end{tabular}
\caption{Same as for table  \ref{wasp43}, but for the stars,   Kepler-91 \citep[see e.g.][]{Lillo}} 
 and HD32518   \citep[see] []{Doll} .
\label{Kepler91}
\end{table}
\begin{table}
\begin{tabular}{|c|c|c|c|c|c|}
\hline
 & $m$ & $R_{pl}$ & $P_{orb}$ & e\\
\hline
 Kepler-91b &0.73$\pm$0.13$M_J$ & 1.384$\pm$0.054$R_J$ & 6.2466 d & 0.066$\pm$0.013 \\
\hline
 HD32518b & $3.04\pm$0.68$M_J$ & N/A & 157.54 d & 0.01$\pm$0.03 \\
\hline
\end{tabular}
\caption{Same as for table \ref{wasp43b}  but for the exoplanets,  Kepler-91b\newline  
 \citep[see e.g.][]{Lillo,Barclay}
 and \newline  
  HD32518b \citep[see][]{Doll}.}
\label{Kepler91b}  
\end{table}
\begin{figure}
\begin{center}
\vspace{8cm}\includegraphics{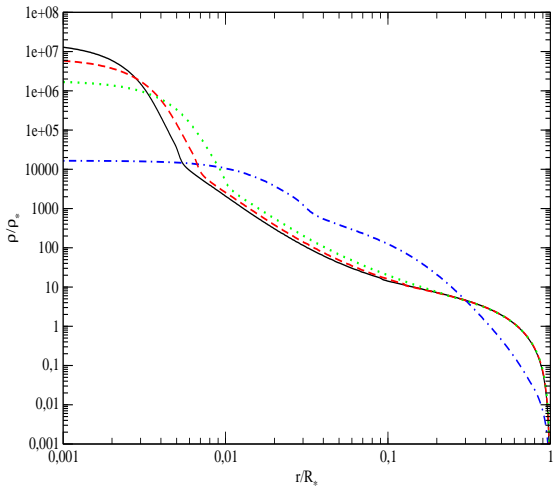}
\end{center}
\vspace{0.5cm} \caption{The dependence of density of the model of Kepler-91 specified in table \ref{Kepler91}
expressed in units of the mean density $\rho_*$ on dimensional radius
$r/R_*$. See the text for a description of  particular curves.} \label{denKep}
\end{figure}
\begin{figure}
\begin{center}
\vspace{8cm}\includegraphics{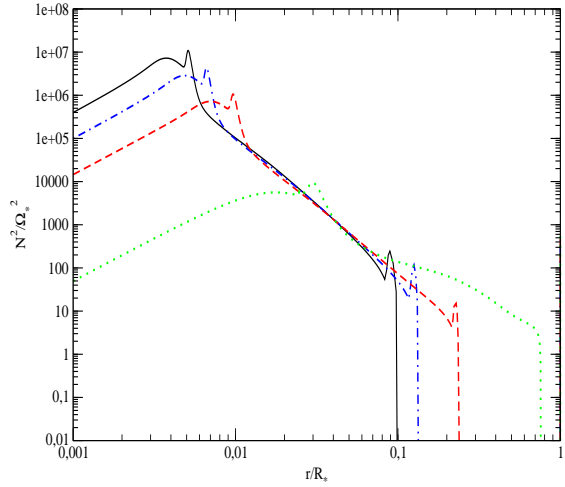}
\end{center}
\vspace{0.5cm} \caption{Same as for Fig. \ref{denKep}, but for the square of the Brunt - V$\ddot {\rm a}$is$\ddot
{\rm a}$l$\ddot {\rm a}$  frequency expressed in units of  $\omega_*^2=GM/ R_*^3$.} \label{N2Kep}
\end{figure}
 We have checked
that  tidal evolution is essentially insignificant for  $T_{age} < 3.5Gyr$.
Accordingly  we consider 14 stellar
models with ages in the range $3.51Gyr < T_{age} < 4.26Gyr$, with the  time interval between them  decreasing
at later times to account for  more rapid  evolution of the star   (see Fig. \ref{RT}). In Figs. \ref{denKep} and
\ref{N2Kep} we show, respectively,  the density and  Brunt - V$\ddot {\rm a}$is$\ddot
{\rm a}$l$\ddot {\rm a}$  frequency as a function of radius.
Solid, dashed, dotted and dot dashed curves correspond to $T_{age}=4.26$,
$4.21$, $4.11$ and $3.7Gyr$, respectively. One can see that at the latest time the stellar structure assumes a
typical red giant form with a  highly centrally condensed  core and an  extended convective envelope.
\begin{figure}
\begin{center}
\vspace{8cm}\includegraphics{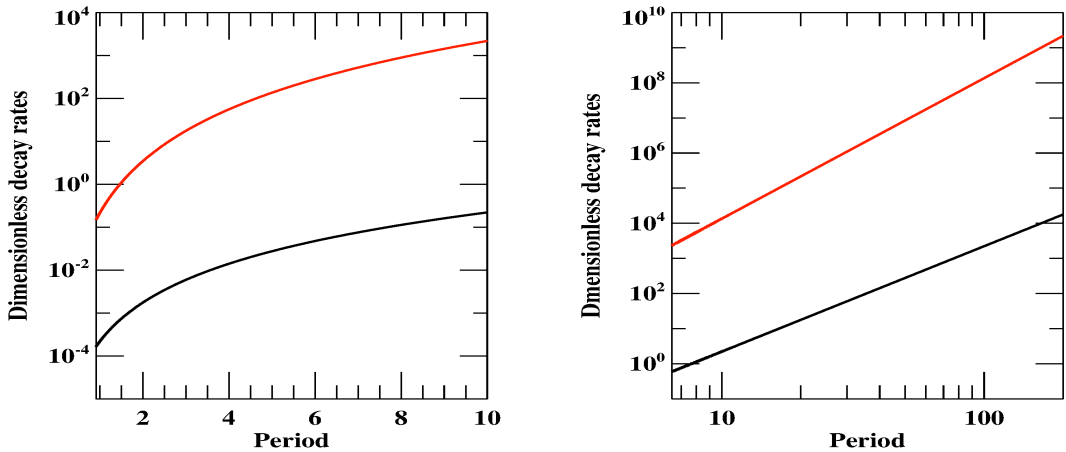}
\end{center}
\vspace{0.5cm} \caption{  The left panel shows  $\gamma/\omega$  ( black curve )and
 $\gamma/(d\omega/dn)$ (red curve) as functions of the orbital period in days (see the caption to Fig. \ref{No}) for the model of  Kepler-91 listed in table \ref{Kepler91}.
 The corresponding plots for the model of  HD32518 also  listed in table \ref{Kepler91} are shown in the right panel.}
  \label{Gia}
\end{figure}
\subsubsection{Properties of normal mode decay rates}\label{NMdecayratesKepler91}
Following the procedure outlined in sections \ref{Modedecay} and \ref{MLDcrit} we evaluated the normal mode decay rates as a function
of forcing frequency, $\omega,$ for the model of Kepler-91 listed in table \ref{Kepler91}.
For reference purposes we also did this for the model of HD32518 also listed in table \ref{Kepler91}.
This is also on the   giant branch.
The ratio of the decay rate of a normal mode to its angular frequency, $\gamma/\omega,$ and
 the ratio of the decay rate of a normal mode to the mode angular frequency interval, $\gamma/(d\omega/dn)$
is shown as a function of $\omega$ for these models in  Fig.\ref{Gia}.

For Kepler-91, $\gamma/\omega < \sim 0.1,$  for periods less than $10$ days 
justifying validity of the
quasi-adiabatic approximation. Note that this quantity $\sim 0.03$ at a  period of $6.25$ days  corresponding to Kepler-91b.
 The quantity  $\gamma/(d\omega/dn) > 1$  for periods exceeding $1.5$ days,  indicating  that  the MLD regime applies for orbital periods characteristic of Hot Jupiters. In the case of HD32518, $\gamma/\omega~<~\sim~1,$ for periods $ < 8$ days  with  $\gamma/(d\omega/dn) >1$ for periods of a few days showing again 
that the MLD
regime 
holds. Note that  for an orbital period of  $157.5$ days  corresponding to HD32518b,  $\gamma/\omega  \sim 10^4$ 
demonstrating a dramatic failure of the quasi-adiabatic
approximation. In this  case $g$ modes are not excited indicating that an  equilibrium tide approach should be followed.

\section{Tidal evolution}\label{Tidalevolution}

\subsection{ Results for Main sequence stars}\label{MSresults}


\begin{figure}
\begin{center}
\vspace{8cm}\includegraphics{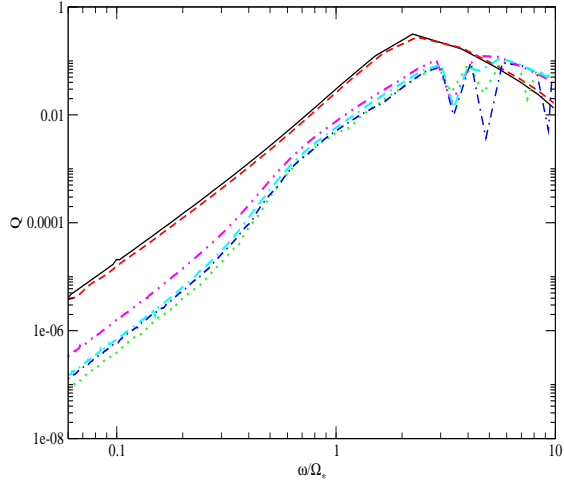}
\end{center}
\vspace{0.5cm} \caption{Overlap integrals as functions of mode  eigenfrequency for the main sequence stellar models.for which
orbital evolution was considered.
Solid, dashed, dotted, dot dashed, dot dot dashed and dot dashed dashed  curves correspond to  WASP-43,  Ogle-tr-113, WASP-12, models A and B, and
WASP-18, models A and B, respectively.} \label{Q}
\end{figure}
In order to calculate the orbital evolution by solving (\ref{t1}) - (\ref{t9})
it is necessary to evaluate the overlap integrals for the stellar model under consideration \citep{Chernov}.
We show overlap integrals, $Q_k$, obtained for our models of main sequence stars
in Fig. \ref{Q}. For all models $Q_k$ is found to  sharply decrease as  $\omega_{j=j(k)} $ decreases.

In case of WASP-43 and  Ogle-tr-113  the shape of the curves is very similar to what is  obtained from  a Solar model, while in  the case
of WASP-12 and WASP-18 the results are   closer to those obtained for  more massive stars \citep[see e.g.][(ChIP)]{ChIP}.
At a given frequency,  $Q_{k}$ is markedly smaller in the latter case as compared to the  former case.
This is attributed to a smaller  relative size of the
 convective envelope in the models of WASP-12 (see ChIP).
\begin{table}
\begin{tabular}{|c|c|c|c|c|c|}
\hline
   & $\omega=2\Omega_{orb}$ & $|d\omega_j/dj||_{j(k)}$  \\
\hline
 WASP-43 & 1.7879e-4 & 1.1086e-5 \\
\hline
 WASP-12A & 1.3326e-4 & 3.7050e-06 \\
\hline
 WASP-12B & 1.3326e-4 & 4.0569e-06 \\
\hline
 Ogle-tr-113 & 1.0153e-4 & 3.4381e-6 \\
\hline
\end{tabular}
\caption{Values of the forcing frequency and the frequency separation between successive  normal modes   in its vicinity
for some  non-rotating stellar models in seconds$^{-1}$.}
\label{frec}
\end{table}

We also give, for a reference, the values of the forcing frequency $\omega=k\Omega_{orb}$   and the frequency difference between two
neighbouring modes for the mode which is closest to  resonance, $|d\omega_j/dj|_{j(k)}$, for the  stellar models considered in detail in table \ref{frec}.
We take $k=2,$ which is the case of interest for our calculations  and
neglect the effect of rotation.
We note that the
decay rate of the mode nearest to resonance  is given by  $\gamma=\kappa |d\omega_j/dj|_{j(k)}$. From table \ref{frec}  we see that, $|d\omega_j/dj|_{j(k)} \ll 2\Omega_{orb}$,
is small corresponding to a dense spectrum of modes, thus justifying the use of formalism developed in IPCh.

\begin{figure}
\begin{center}
\vspace{8cm}\includegraphics{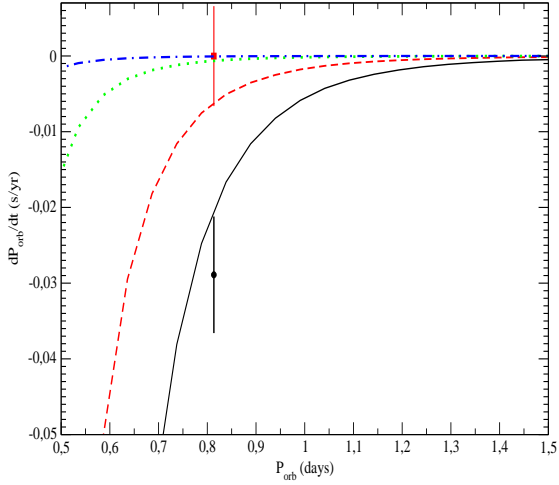}
\end{center}
\vspace{0.5cm} \caption{Results related to our model of WASP-43 are shown. Different curves represent the time derivative of orbital period, $\dot P_{orb}$, in units of $s/yr$ as functions of orbital period in days, for different values of the quantity $\kappa$ parametrising mode dissipation rate. See the text for a description
of particular curves.  The black circle and  red square show the positions of two proposed values of $\dot P_{orb}$ inferred
from analysis of observational data by \protect\cite{Jiang2016} and \protect\cite{HoyerAJ}, respectively.} \label{Pwasp43}
\end{figure}

\begin{figure}
\begin{center}
\vspace{8cm}\includegraphics{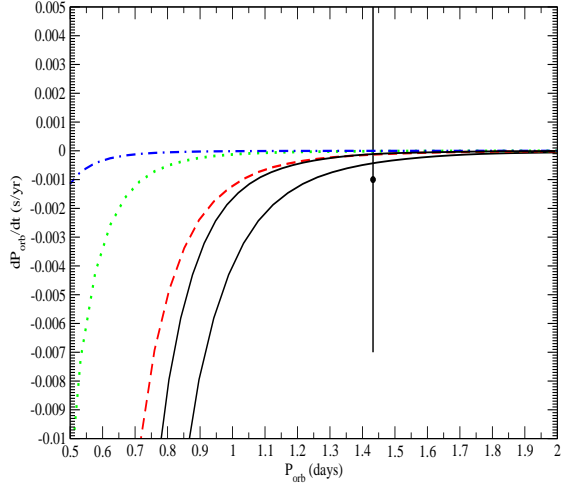}
\end{center}
\vspace{0.5cm} \caption{Same as Fig. \ref{Pwasp43}, but for the star  Ogle-tr-113. Note that in this case there are two solid curves corresponding to
the MLD regime, the curve taking on  smaller (larger) values  at a
given $P_{orb}$ corresponds to the non-rotating (rotating) star.
The  value of $\dot P_{orb}$ obtained from observations
 and its error bar are taken from \protect\cite{HoyerMNRAS2016}. } \label{Pogle}
\end{figure}

\begin{figure}
\begin{center}
\vspace{8cm}\includegraphics{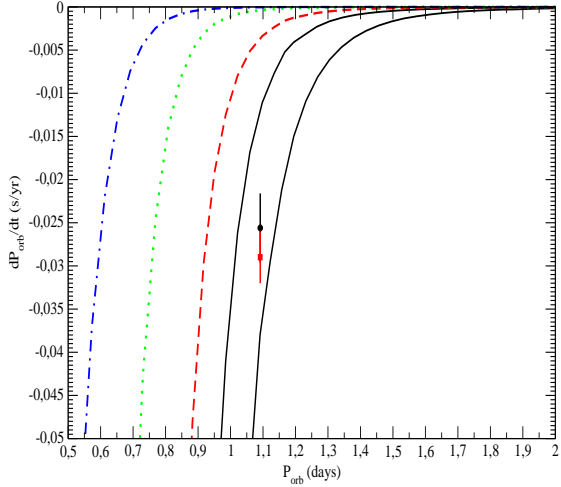}
\end{center}
\vspace{0.5cm} \caption{Same as Fig. \ref{Pwasp43}, but for the star  WASP-12. The solid curve with smaller (larger), $|\dot P_{orb}|$,
corresponds  to  model A  (B).   The other curves correspond to model A. The indicated positions of the value of $\dot P_{orb}$ obtained from observations and their error bars
are taken from \protect\cite{Maciejewski2016} and \protect\cite{patra}, with smaller and larger absolute values of
$\dot P_{orb}$ corresponding to the former and latter references, respectively.
} \label{Pwasp12}
\end{figure}

\begin{figure}
\begin{center}
\vspace{8cm}\includegraphics{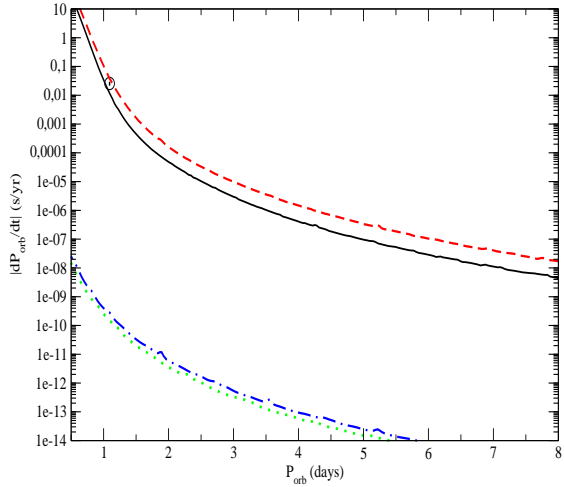}
\end{center}
\vspace{0.5cm} \caption{$|\dot P_{orb}|$ as a function of orbital period.
Solid  and dashed curves,  obtained
with the use of our formalism,  respectively  correspond to models A and B  for WASP-12. The  dotted and dot dashed curves are calculated according to the Zahn prescription, applied to
models A and B, respectively. All curves have been calculated in the assumption that the MLD regime applies.
} 
\label{Pwasp12log}
\end{figure}

\begin{figure}
\begin{center}
\vspace{8cm}\includegraphics{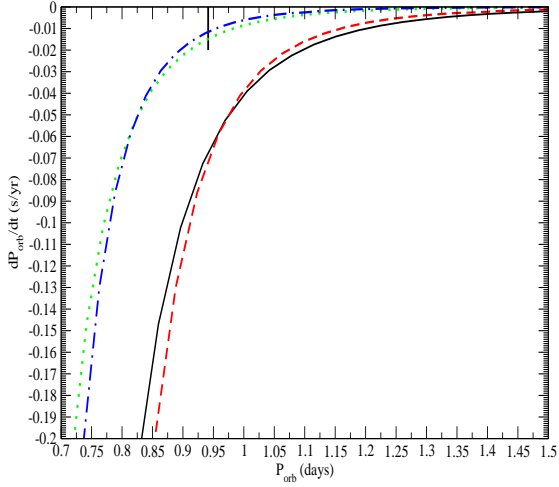}
\end{center}
\vspace{0.5cm} \caption{$\dot P_{orb}$ as a function of orbital period 
calculated for the system WASP-18. The vertical solid line indicates the published
observational limits on the orbital evolution of the system. All theoretical curves are calculated assuming the MLD regime. Solid and dashed
lines show the theoretical results for models A and B and assuming a non-rotating star  respectively  The dotted and dashed lines are
respectively  for models A and B but with the star assumed to have  $P_r=5.39d.$ } \label{Pwasp18}
\end{figure}

We use these overlap integrals  in
 equations (\ref{t1})-(\ref{t9})
 to  enable the orbital evolution to be calculated together
 with the time derivative of orbital period, $\dot P_{orb}.$
 Results of the calculations of, $\dot P_{orb},$
together with data inferred from observations,  are shown in Figs. \ref{Pwasp43}, \ref{Pogle} and \ref{Pwasp12},
for the models of WASP-43b,  Ogle-tr-113b and WASP-12b, respectively.  Solid, dashed, dotted and dot dashed curves respectively
correspond to a formally infinite value of $\kappa $  making  $D=1,$  implying the MLD regime,
$\kappa=0.1$, $0.01$ and $0.001.$
We have checked that the result corresponding to $\kappa = 1$ gives a curve almost indistinguishable from the solid one.
In  addition, for values of $\kappa < 0.001,$ $\dot P_{orb}\propto \kappa,$ as follows from equation (\ref{smallkap}).

Two different values of $\dot P_{orb}$
obtained from analysis of observational data are indicated in  Fig. \ref{Pwasp43}.
A relatively
large value of  $|\dot P_{orb}|\sim 0.03 s/yr$ was reported by \cite{Jiang2016}, whereas \cite{HoyerAJ} claim that $|\dot P_{orb}|$ is significantly  smaller
and consistent with zero.
 As seen from Fig.
 \ref{Pwasp43} the larger value of $|\dot P_{orb}|$ can be easily explained by  as resulting from  tidal evolution in the
MLD regime. On the other hand if the result
 of  \cite{HoyerAJ}  holds, the
action of tides in this systems is very much  weaker than that predicted within the framework of that  regime.

In the limit of weak dissipation, when the system is mid way between resonances,
the evolution rates, including, $|\dot P_{orb}|,$ have to be multiplied by a factor
 $\pi\kappa = \pi\gamma /( |d\omega_j/dj|_{j= j(k)})$
(see equation ({\ref{WEAKR})  and discussion above).  For WASP-43 we find that this quantity is $\sim 4\times 10^{-8}$
resulting in an extremely small $|\dot P_{orb}| \sim 10^{-9} s/yr.$
Thus, the observational result of   \cite{HoyerAJ}  is consistent with tides being in the linear regime with radiative damping operating
in the weakly dissipative limit.
\footnote{In
this connection we remark that
much larger values of $|\dot P_{orb}|$ could be formally obtained by bringing the system closer to resonance
but non linear effects should then be considered.}

In the case of Ogle-tr-113b,  which is a relatively fast rotator,  we
show $\dot P_{orb}$ calculated in the MLD regime for a non-rotating star, and for a star with rotational period $P_r=7.79d$  in Fig. \ref{Pogle}, with
solid curves taking on respectively  smaller and larger values  at a given $P_{orb}.$  Note
that in the latter case we assume that the resonant frequency is expressed
in terms of the orbital frequency and the angular frequency of rotation,
$\Omega_r=2\pi/P_r$, through  $\omega=2(\Omega -\Omega_r)$.

In the case of Ogle-tr-113b,  Fig. \ref{Pogle} 
shows that
the value of  $|\dot P_{orb}|,$ obtained from observations
is consistent with the system operating in
 the MLD regime, with the mean value
 being very close to the theoretical curve for the non-rotating case. However,  the reported observational
errors are so large that even positive values of $\dot P_{orb}$  are not excluded. Accordingly, no definite conclusion about the regime in which tides operate can be made in this case.

The results presented in  Fig. \ref{Pwasp12} for WASP-12
indicate that the MLD regime is fully compatible with the available data. Model A gives a slightly smaller value of $|\dot P_{obs}|$ than that
is obtained from observation, while model B gives a slightly larger value. It is evident that one could
obtain a perfect agreement using a model of WASP-12 with parameters intermediate to those have been
employed in models A and B.

Since WASP-12 is a star possessing  a convective core, albeit a small one,  as a matter of interest we can apply the theory of \citet{Z1,Z2} to this star and compare  results.
 This comparison is shown in Fig. \ref{Pwasp12log}, where we plot absolute values of $\dot P_{orb}$ calculated
in the framework of our formalism,  under the assumption of the MLD regime,  as solid and dashed lines and in the Zahn theory as dotted and dot dashed lines, for
models A and B, respectively. One can see that our approach  gives  very much larger tidal evolution rates,
being  $\sim 10^{9}$ times larger than
given by the the Zahn theory.
This result is, however, expected since
this 
discrepancy arises because the Zahn theory is based on the modes being excited at the outer boundary of convective core, 
which has small radius
(see Fig. \ref{N2} ), whereas
in our case the important region is near the inner 
boundary of the convective envelope, 
see also \cite{Goodman}.

Finally, let us discuss the system WASP-18. For this system we present
theoretical dependencies of $\dot P_{obs}$ on $P_{obs}$ in Fig. \ref{Pwasp18}.
All curves are obtained assuming that the MLD regime operates.
We see that the results obtained  for the non-rotating star are
clearly incompatible with the observational limits. However, when the
effect of stellar rotation is taken into account both models considered
are within the limits. Interestingly, these models predict $|\dot P_{orb}|
\approx 0.01-0.015 s/yr$, which is just slightly smaller that the published
upper bound  $|\dot P_{orb}| < 0.02 s/yr$. Thus, observations
should allow us to either confirm or discard the
possibility that  tidal interactions occur in the MLD regime in this system in the near future.

In summary, the limited observations available suggest that main sequence stars orbited by close giant companions could sometimes be in the MLD
regime and  sometimes not. The latter case is what would be expected from the linear theory of tides for which radiative damping operates.
In the former case non-linearity needs to be invoked in order to provide adequate dissipation \citep[see eg.][]{Barker}.
It is clear that more observations are needed in order to make a robust conclusion about  whether and how often   tidal evolution operates in the MLD regime
 in systems containing main sequence stars
and close-in Hot Jupiters.

\subsection{Kepler-91b} \label{SKepler91b}

We now  consider the  tidal evolution of  Kepler-91b.
 As can be seen  from Figs. \ref{denKep} and
\ref{N2Kep}  the star Kepler-91 is a red giant with an extended convective envelope.
This  means that in addition to $g$ modes being excited by  tides there could be  other significant factors influencing orbital
evolution including effects due to quasi-static tides, as well as effects associated with a powerful stellar wind, namely  stellar mass loss, mass accretion
by the planet and hydrodynamic drag exerted on the planet by the wind.
 To take account of these, we write  the  equations
governing  orbital evolution in the form
\begin{equation}
{\dot a\over a}=-{2\over T_{a}}-{2 \over T_{a, QSt}}+{\left({\dot a\over a}\right)}_{W} \quad {\rm and}  \quad {\dot e\over e}=-{1\over
T_{e}}-{1\over T_{e, QSt}}+{\left({\dot e\over e}\right)}_{W},
\label{k1}
\end{equation}
where $T_a$ and $T_e$ are given by equations (\ref{t11}) - (\ref{t9}) and the
quantities subscripted with, $QSt,$ and, $W,$ represent rates of  change of the orbital parameters due to
the action of quasi-static tides and  the stellar wind, respectively.  
{  We note that as discussed in Section \ref{NMdecayratesKepler91}, the MLD regime is expected to apply
for the current model of Kepler-91 and the  orbital periods we have  considered. 
Thus  standing waves and resonant modes cannot be set up 
(see Section \ref{Tidev}).
 This means that resonance locking
 is   not expected to be occurring  at the present evolutionary stage.}
We go on to  discuss {   quasi-static tides and  the stellar wind}  in turn.

\subsubsection{Quasi-static tides}\label{Quasi-static tides}
We follow \citet{Z2} and assume that tidal dissipation takes place as a  result of turbulent viscosity
acting on the quasi-static equilibrium tide in the convective  envelope.
We treat quasi-static tides in the simplest possible approximation using the results of  \cite{Z3}, namely,
we use  equation (17)  of that paper for the evolution of semi-major axis and equation (18) of that paper for the evolution of eccentricity. 
We set the  angular velocity of stellar rotation and  the orbital eccentricity to zero in the right hand side  of these equations.
 The characteristic time scales  of orbital evolution due to quasi-static tides,
 $T_{a, QSt}$ and $T_{e, QSt}$, are shown in Appendix as euqations (\ref{A1}) and (\ref{A2}), respectively.
They depend on parameters $\lambda_{l}$ defined in equations (\ref{A3}) and (\ref{A4}),  for  
$l=1,2,3,$, where $l$ are numbers of Fourier harmonics in the 
decompostion of the perturbing potential in Fourier series in time.
They 
in  turn depend on  the quantities $\eta_{l}=2lt_{f}/ P_{orb}.$ 
Here $t_{f}$ is the characteristic
time of turnover of convective eddies defined as 
\begin{equation}
t_{f}=\left ({M_*R_*^2}\over {L}\right)^{1/3}.
\label{qs1}
\end{equation}

The weakening of turbulent viscosity
in the regime of $\eta_{l} \gg 1$ as  noted by  eg. \cite{GoldN}, results in reduction of $\lambda_{l}$ with increasing  $\eta_l$.
The form of the dependence of $\lambda_{l}$ on $\eta_l$  that should be used is unknown. In the past a power-law dependence
with either $\lambda_l \propto \eta_l^{-2}$  \citep[eg.][]{GoldN} or $\lambda_l \propto \eta_l^{-1}$ \citep[ e.g.][]{Z3} has been assumed.
However, the actual situation may be more complex with the effective viscosity even being negative in some cases
\citep[see][]{Ogilvie2012}.
Here, for simplicity,  we shall   treat this dependency in the simplest possible way.
 Namely, when $\eta_l < 1$ we use
equation (13) of \citet{Z3}  when calculating $\lambda_l$.  Then  these quantities are found  not depend on $\eta_l$ and $l$ so we may write
$\lambda_l\equiv \lambda .$
When $\eta_l > 1$  we adopt  $\lambda_l=\lambda /\eta_l.$

In order to  calculate $\lambda$ we use parameters such as radius of the base of convective envelope, etc.  for  a  grid of models with different ages,  that is
used for our calculation of the effect of  dynamic tides on the orbit.
We find $T_{a, QSt}$ and $T_{e, QSt}$ for these models and use linear interpolation  to obtain them
for intermediate  ages  of the star. In following the procedure outlined above, we stress the considerable uncertainty in estimating
 the effects of turbulence acting on quasi-static tides and that the effective turbulent viscosity and consequent effects on
 orbital evolution may be significantly overestimated.

\subsubsection{Stellar wind}\label{Stellar wind}
In order to 
evaluate the rate of mass loss from stellar wind, $\dot M_{W}$, we use the Reimers law \citep{R}
\begin{equation}
\dot M_{W}=4\cdot 10^{-13}\eta_{R}\left({L\over L_{\odot}}\right)\left({M\over M_{\odot}}\right)
\left( {R_{\odot}\over R_*}\right)
M_{\odot}/yr^{-1},
\label{w1}
\end{equation}
where $L$ is the stellar luminosity.
 From the conservation of angular momentum it follows that the  semi-major axis
will evolve as a result of the mass loss according to
\begin{equation}
\left({\dot a\over a}\right)_{W}={\dot M_{W}\over M}.
\label{w2}
\end{equation}
Thus it will take on larger values as  a result of a positive  rate of mass loss.

One can see that the change of the gravitational field of the star due to mass loss does not
lead to an appreciable change of orbital
eccentricity since both the eccentricity
and the angular momentum  are  adiabatic invariants when the orbit changes
on account of the changing mass of the central star.
Accordingly, we set $\dot e_{W}=0$.

\subsubsection{Hydrodynamic drag}\label{hydrodynamic drag}
Let us estimate the effect of hydrodynamical drag  exerted on the planet.
We  first  calculate the ratio
of Bondi-Hoyle radius $R_{BH}=2Gm/(v_k^2+c^2_s)$, where $v_k$ is Keplerian velocity
taken to be  $v_k=\sqrt{GM/ a}$ for a near circular orbit,
 and $c_s$  the sound speed of the wind,
to the planet radius $R_{pl}.$
 This is  easily done with help
of tables  \ref{Kepler91} and  \ref{Kepler91b}
together with the  assumption  that $c_s \sim 30km/s \ll v_k,$
with the result that $R_{BH}/R_{pl}\sim 0.16$\footnote{The ratio of the  gravitational drag force to the hydrodynamic  drag force  is proportional to the product of the square
of  $R_{BH}/R_{pl}$ and the usual Coulomb
logarithm, see e.g. \cite{Th}. Assuming that the largest and the smallest scales in the problem are  respectively  the semi-major axis  and $R_{pl},$ which have a  ratio  $\sim 100$,
  the ratio of two the drag forces is found to be of order $\sim 0.1$.}.

This means that gas trajectories are not significantly
deflected by  the planet's gravitational field  before meeting the planet.
Thus, to make
 a rough estimate of the rate of energy exchange with the planetary orbit  per unit of time,
 $\dot E_{HD}$, we  can simply calculate the rate of energy flow due to
  gas elements moving
 with Keplerian velocity through a target with cross-section equal to $\pi R_{pl}^2$. This gives
\begin{equation}
\dot E_{HD}\sim -{\pi \over 2}\rho R_{pl}^{2}\left({GM\over a}\right)^{3/2}= -{1\over 8}\left({R_{pl}\over a}\right)^2 \left({GM\over a}\right)^{3/2}{1\over v_{W}}\dot M_{W},
\label{w3}
\end{equation}
where the negative sign  occurs because hydrodynamic drag decreases the  orbital energy,
 $\rho $ is the wind  density
and we have used  the law of mass conservation for the wind in the form
 $\rho~=~\dot M_{W}/(4\pi a^2 v_{W}).$
 The  rate of change of orbital semi-major axis is then found from  setting
\begin{equation}
 \dot E_{HD}=\frac{GmM }{ 2a^2}\dot a_{HD},
\label{Eadot}\end{equation}
 where $\dot a_{HD}$ is the rate of change of
 semi-major axis due to hydrodynamic
drag. Combining (\ref{Eadot})
 (\ref{w3}) and  (\ref{Eadot})
we obtain
\begin{equation}
\left({\dot a\over a}\right)_{HD}\sim -{1\over 4}\left({R_{pl}\over a}\right)^2 {v_K\over v_{W}}{\dot M_{W} \over m}.
\label{w4}
\end{equation}
It is convenient to express $({\dot a / a})_{HD}$ in terms of
 $({\dot a /  a})_{W}$ by writing
 $({\dot a/ a})_{HD}=-f({\dot a /  a})_{W}$,
where the explicit form of the parameter $f$ follows from (\ref{w2}) and (\ref{w4}) as
\begin{equation}
f = {1\over 4}\left({R_{pl}\over a}\right)^2 {v_K\over v_{W}}{M \over m}
 \approx 0.1\left({30km/s\over v_{W}}\right),
\label{w5}
\end{equation}
where we have used the parameters given in  table \ref{Kepler91b}
 in order to obtain the last equality.

Equation (\ref{w5}) 
tells us  that  effects due to hydrodynamic drag
are expected to be smaller than those associated with the change of mass of
the star unless the  wind velocity is
unrealistically small $v_{W} < 3km/s$.
Therefore, we shall neglect hydrodynamic drag  in our analysis of the orbital evolution
of Kepler-91b.

\subsubsection{Orbital evolution}\label{Orbital evolution}

\begin{figure}
\begin{center}
\vspace{8cm}\includegraphics{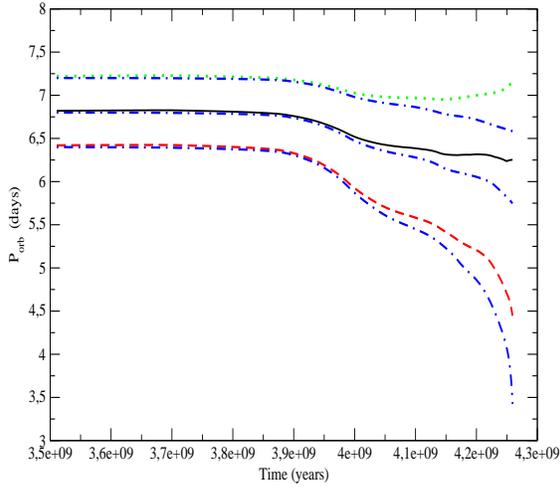}
\end{center}
\vspace{0.5cm} \caption{The evolution of orbital period obtained using  our model
 of the tidal interaction of Kepler-91b. Solid, dashed and dotted
curves correspond to different initial orbital periods taken
when the age of the star was  $T_{in}=3.5Gyr$ and evolution of the system was derived from
equation (\ref{k1}). The  dot dashed curves are for the same initial periods,
 but with the influence of effects associated
with the stellar wind and quasi-static tides being neglected.} \label{EvKep1}
\end{figure}

\begin{figure}
\begin{center}
\vspace{8cm}\includegraphics{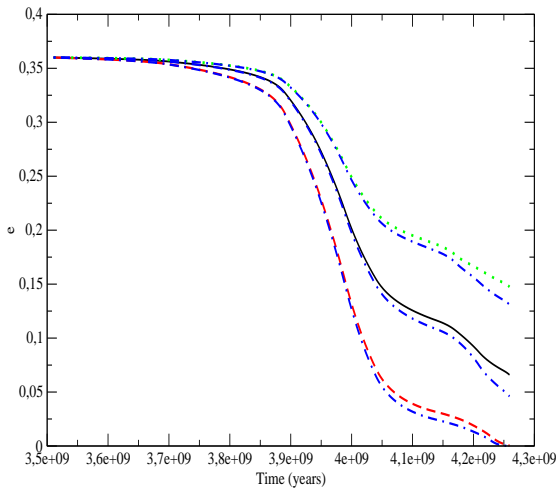}
\end{center}
\vspace{0.5cm} \caption{Same as Fig. \ref{EvKep1}, but for the evolution of eccentricity with time. The initial eccentricity
$e_{in}$ was taken to be equal to $0.36$ in  all cases.} \label{EvKep2}
\end{figure}
As already stated  above we calculated  $T_a$ and $T_e$ for the set of models of Kepler-91 with different ages and linearly
interpolated  them to be able to find these time scales at an arbitrary  intermediate age.
 Then, we integrated
equation (\ref{k1}) numerically, taking into account,  in addition to the excitation of $g$ modes in the MLD regime, the effect of quasi-static tides in the approximation specified above, and
the effect of changing gravitational field of the star due to mass loss.
The results for the evolution of orbital period and eccentricity are  respectively shown in Fig. \ref{EvKep1}
and Fig.\ref{EvKep2}.

 Our evolutionary  tracks depend on the
values of the orbital period and eccentricity adopted at the  initial time taken to be when the age of the star was $T_{in}=3.5Gyr.$
These were  chosen in such a way that
the solid curves  in Figs. \ref{EvKep1} and \ref{EvKep2}  reproduce the observed values, $P_{orb}(T_{fin})\approx 6.25d$
and $e(T_{fin})\approx 0.066$ at final time when the age of the star was $T_{fin}=4.26Gyr$  as currently estimated (see table \ref{Kepler91}).
 This requires $P_{orb}(T_{in})\approx 6.8d$ and $e(T_{in})\approx 0.36$.
 Thus  tidal evolution
can diminish a rather large value of initial eccentricity to the small observed value, while the value of orbital
period  changes by less than 10 per cent.
 Dashed and dotted curves show the results corresponding to
slightly smaller and slightly larger initial orbital periods, $P_{orb}(T_{in})\approx 6.4d$ and $P_{orb}(T_{in})\approx 7.2d$,
respectively.
One can see from Figs. \ref{EvKep1} and \ref{EvKep2}  that the evolution looks rather different for these cases as compared to the case fitting
the data. While the case with smaller initial period shows a violent tidal evolution at late times leading to a
strong decrease of orbital period, the case with larger initial period  is such that the period increases   at late times due to the
effect of mass loss from the star.
This feature could lead to an understanding of  the  present day orbital parameters of Kepler-91b, since
 tidal and mass loss effects  on the evolution of  the semi-major axis can be  nearly balanced, while
at the same time,  tidal effects lead to  significant  decrease of eccentricity.
Note that  the results plotted in Figs. \ref{EvKep1} and \ref{EvKep2}   show that quasi-static tides considered
in our approximation, which we argued are likely to be overestimated,  do not  significantly influence the evolution of the system.

\begin{figure}
\begin{center}
\vspace{8cm}\includegraphics{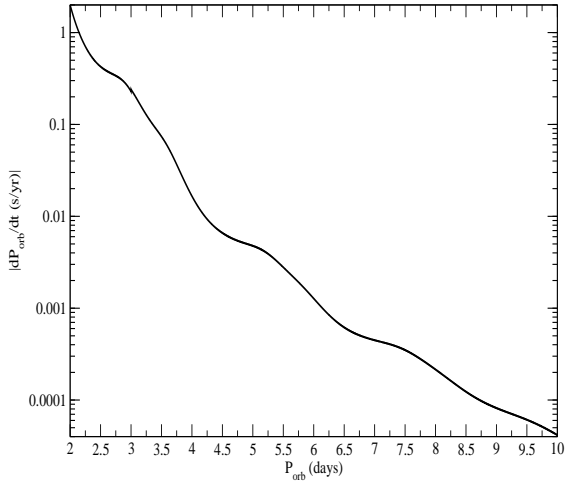}
\end{center}
\vspace{0.5cm} \caption{ $|\dot P_{orb}|$ as a function of orbital period
for a star with parameters of the present day model of Kepler-91 and a planet with the mass of Kepler-91b.} \label{EvKep3}
\end{figure}
Finally,  for reference and comparison with the main sequence models,  we show the dependence of the absolute value of $\dot P_{orb}$ on the orbital period for
a system containing a star with parameters appropriate to the present day state of Kepler-91 and a planet with  the mass
of Kepler-91b. Only dynamic tides  in the MLD regime are taken into account in this calculation. One can see that at orbital periods of
$'\sim 6d$ corresponding to Kepler-91b,  $|\dot P_{orb}|$ is rather small,  being of the order of $0.001s/yr$. Note too  that the effect
of mass loss would reduce it still further.

\section{Discussion and Conclusions}\label{Discuss}
In this Paper we have  applied   our general formalism  for determining effects due to
dynamical  tides  developed in \cite{IPCh} (IPCh)  to calculate the expected orbital evolution in  observed systems containing a Hot Jupiter.
Our formalism is based on the normal mode approach to the problem, it contains within it the well known Zahn theory of
dynamical tides. Contrary to the Zahn theory, which applies only in the asymptotic limit of small
tidal forcing  frequencies  and only for {   hot stars with convective
cores and radiative envelopes }     \citep[see][]{Z1,Z2},
our formalism allows one to consider more general and realistic stellar models together with
 forcing frequencies that are not asymptotically small.
 We pay  special attention to the question of whether or not
the assumption that the propagation time of wave trains excited by tides through propagation zones,  is longer  than their dissipation time, corresponding
to the regime of moderately large {  dissipation} (MLD),  is supported by the analysis of present day observations.
We note in passing that  the Zahn theory is applicable in this regime also.

We consider several main sequence (MS) stars with Hot Jupiter companions,
for which either the rate of change of orbital period, $\dot P_{orb}$, or an upper limit for it have been reported,
as well as  the evolved star Kepler-91,
 which has a Jupiter mass companion, Kepler-91b, on a close in orbit.
We demonstrate that although the linear mechanism of radiative damping  of tidally excited modes
is not effective enough to justify the assumption of the MLD regime  for  the  systems containing a main sequence star,
it results in  the  Kepler-91 system  evolving in the MLD regime.
{  We  recall that in cases for which the MLD regime does not  operate,  relatively weak
dissipation is implied unless there is a  resonance with a normal mode. }
\subsection{Systems with stars {  on} the main sequence}
The systems containing MS stars and Hot Jupiters we considered  in Section \ref{MSresults} were  WASP-43,  Ogle-tr-113, WASP-12 and WASP-18.
The former two  contain  Sun-like stars with radiative interiors and convective envelopes, while the latter two
are relatively more massive and have the complimentary  structure with convective cores and  envelopes {  which are for the most part radiative with  there being a relatively small convective region near the surface}.
{  We remark that  on account of their low mass and the   mismatch between the convective turn over time and the inverse tidal forcing frequency,  tidal dissipation in these envelopes is expected to 
be ineffective \citep[see eg.][]{Barker}.}
In addition, the stars,  Ogle-tr-113 and WASP-18 appear to be rotating sufficiently  rapidly that the effect of rotation
should to be taken into
account when calculating $\dot P_{orb}$.
Note that this  weakens the  tidal interaction when the  angular momentum vectors associated with  stellar rotation and
the orbital motion are  aligned as has been assumed in this paper.

Values of $\dot P_{obs}$ have
been  obtained  for  WASP-43   by  \citet{Jiang2016} and \citet{HoyerAJ}.
Although the former authors give
an absolute value of $\dot P_{orb}$, which is consistent  the assumption of MLD regime,
the value given by the
latter authors is too small to be consistent with it.
 WASP-43 appears to be a slow rotator, and, therefore, the inconsistency
of  this assumption with  the analysis of \citet{HoyerAJ} cannot be alleviated by taking rotation into account. On the other hand it is consistent with being in a weak dissipation regime and the
tidal forcing frequency being mid way between neighbouring normal mode frequencies.

In case of Ogle-tr-113,  models with tides operating  in the MLD regime
with and without rotation   are consistent  with the present
 observations.  However,  error bars are so large  that even positive values of $\dot P_{obs}$
are not necessarily  excluded.

WASP-12 appears to   show  the best consistency  with the assumption of   being in the MLD regime.
Errors bars are small enough in this case to accommodate   results provided by both of our models for
this star. Note, however, that \cite{patra} also consider another scenario for the observed changes in occulation
times base on apsidal precession giving it, however, less probability that the one based on the orbital decay
due to tides.
It is of interest to note that a direct application of the  Zahn theory  to this system gives
values of $\dot P_{obs}$ that are
orders of magnitude smaller ( see Fig. \ref{Pwasp12log}) .

Finally, in the case of WASP-18 only an upper limit  given by  $|\dot P_{obs}| < 0.02s/yr$
is inferred from the  lower limit on the modified tidal quality factor, $Q^{'} > 10^6,$
recently published by  \cite{wilkins}. Although this limit certainly excludes MLD tides operating in non-rotating stars, MLD tides in the star rotating with rotational period $\sim 5.4d$ are in marginal agreement with the published bound on $|\dot P_{obs}|$ giving $|\dot P_{obs}|\approx 0.01-0.015s/yr,$ depending on the stellar model.
   Thus in case of this system, a relatively minor reduction in the magnitude of the observational error,   could  either confirm or exclude the MLD regime.

As we have 
discussed in Section \ref{NMdecayratesMS},  the assumption of the MLD regime,  cannot be justified  assuming the theory of linear damping
of tidally excited modes 
due to radiative  diffusion  applies to the considered MS stars.
Thus, some mechanism of non-linear mode damping must be invoked to account for  the possibility of this regime applying for these objects.
Since both  WASP-43 and  Ogle-tr-113 are Sun-like we can check whether or not they  satisfy the criterion for  wave breaking in their radiative cores given by \cite{Barker} .
We found out that this criterion is not satisfied,  with characteristic mode amplitudes near the centre being several orders of magnitude smaller than needed.
This is explained by relatively young ages of  WASP-43 and  Ogle-tr-113.
As discussed in  \cite{Barker} the wave amplitude near the  centre of a  MS star is proportional to
a positive power of the gradient of the  Brunt - V$\ddot {\rm a}$is$\ddot {\rm a}$l$\ddot {\rm a}$ frequency, and this is relatively small for young MS stars.
A possible non-linear mechanism that could work in the objects with
the considered parameters is  mode decay through non-linear interactions that produce  a large number of 'daughter' modes.
This   was recently discussed by \cite{essick}, who indeed found that it can operate in systems with
Hot Jupiters with periods of the order of one day.

In summarising   our results for  MS stars,  we would like to stress  that the possibility of the  MLD regime operating  generically
in systems containing Hot Jupiters with periods of the order of one day,
has not been definitively  established  at the present time.
When   theory and observation  are compared,
it seems that all except WASP-43 are consistent with being in the MLD regime. However, observational error bars are  large so this may not in fact be  the case.
However  in the case of  WASP-18,  observational bounds on $|\dot P_{orb}|$ are very close to theoretically predicted values,  so that  a modest improvement
in the  former   could  provide  important 
clarification on this issue.

\subsection{Evolved stars}
We considered  the  case of an evolved star with  a close in companion of
 approximately  one Jupiter mass, Kepler-91, in Section \ref{Kepler91b}.
 For this star we have shown in Section \ref{NMdecayratesKepler91}, the existence of linear radiative damping implies that the  MLD regime holds.
This is because of the  dense normal mode spectra found for the  models of this star, in the frequency ranges of interest leading to a resonant mode of very high order.
Accordingly,  dynamical tides  associated with $g$ modes of very high order
are expected to  play a role in the  orbital evolution of this system.

We found that the time scale for orbital evolution  induced by these tides becomes comparable to the  time scale for  evolution of the star during the later stages.
Therefore, stellar evolution must be fully incorporated in the calculations of the evolution of the orbit which will be significant over the lifetime of the star.
We found that  quasi-static tides appear to give only a minor contribution, at least within  the framework of the  simple model adopted.
This assumed  that the effective turbulent viscosity is reduced by the ratio of tidal forcing period to the characteristic turn over time of convective eddies, when this ratio is less  than unity.   Note that it has been argued that there should be  a quadratic
   reduction in the efficiency of  turbulent viscosity  in this regime (\citep[eg.][]{vill, GoldN} (see Section \ref{Quasi-static tides}).
    In that case orbital evolution due to quasi-static tides would be completely  negligible for this system.
   The effect of stellar wind could have been significant  for a planet, with  present day orbital period slightly larger than that of Kepler-91b,  through the effect of mass loss from the system.
   On the other   hand, the effects of   hydrodynamic drag exerted  on the planet by the wind and gas accretion onto the planet  appear to play only a minor role.
   Note, however, that these effects have been considered having adopted a  procedure  which may have been oversimplified (see Section \ref{Stellar wind}).
    This deserves  further investigation as well as a possibly influence of mass loss from the planet due to e.g. its heating by the star.

  We find  the following  orbital evolution  is likely to have taken place during the lifetime
   of Kepler-91b (see  Section \ref{Orbital evolution}).
  It starts to  become significant when the star is approximately 3.5 Gyr old with the orbital period $P_{orb}\sim 6.8d$ being only slightly larger than the current value $P_{orb}\sim 6.2d,$ and the
  eccentricity being rather large  $\sim 0.35$.
  Since the  time scale  for the evolution of the eccentricity
  is shorter than that for the evolution of the  semi-major axis,  the eccentricity relaxes to its present day small value $\sim 0.066$,
  while the orbital period changes only by a small amount.

  We remark that this circularisation occurs independently of effects due to tides raised on the planet which are also expected to lead to
  circularisation. However, in this context we note that the current orbital period of Kepler-91b is large enough  that such tides may not have operated
  significantly during the life time of the star \citep[see eg.][]{IP2007, IP2010}.
  Note that for slightly larger and slightly smaller initial orbital periods the evolution would be qualitatively different.
  In the former case the orbital period actually increases during the  late stages of stellar evolution
  due to the effect of mass loss from the system. In the latter case dynamical tides are very efficient and the orbit rapidly shrinks
  during the later evolutionary stages. Thus,  the orbital parameters of the present day Kepler-91b are rather special,
  since within the framework of our model, only for such parameters do we expect efficient orbital  circularisation, while strong  prior evolution of the  semi-major axis is not expected.

\section*{Acknowledgments}
We are grateful to G. I. Ogilvie for his important remarks and suggestions.

S. V. Chernov and P. B. Ivanov were supported in part by RFBR grants 15-02-08476 and 16-02-01043, by programme 7 of the Presidium of Russian Academy of Sciences and also by Grant of the President of the Russian Federation for Support of the Leading Scientific Schools NSh-6595.2016.2.

\begin{appendix}

\section{Time scales of tidal evolution due to quasi-static tides}

Here we show, for completeness, the time scale for  the evolution of the  semimajor axis, $T_{a,QS}$, and the  eccentricity,
$T_{e,QS}$, due to quasi-static tides, following the paper of  \cite{Z3}. They have the form
\begin{equation}
T_{a,QS}={1\over 6\lambda_{2}}\left({M_*\over m}\right){\left({a\over R_*}\right)}^8t_{f}
\label{A1}
\end{equation}
and
\begin{equation}
T_{e,QS}={1\over 3}{\left({5\over 8}\lambda_1-\lambda_2+{49\over 8}\lambda_3\right)}^{-1}\left({M_*\over m}\right)
{\left({a\over R_*}\right)}^8t_{f},
\label{A2}
\end{equation}
where
\begin{equation}
\lambda_{l}={1\over \max{(1,\eta_l)}}\lambda
\label{A3}
\end{equation}
and
\begin{equation}
\lambda=0.8725E^{2/5}\int^{1}_{x_c}x^{22/3}(1-x)^2dx.
\label{A4}
\end{equation}
The dimensionless radius is  $x=r/R_*$,
  with $x_c$ being  its vallue corresponding  to the inner boundary of the convective envelope. Note 
that we set 
the mixing length parameter $\alpha^{'}$ as defined  in \cite{Z3} to  be unity  in (\ref{A4}). 
The factor $E$ entering (\ref{A4}) is
obtained by matching the density $\rho_c$ at the convective envelope 
 boundary to that obtained from  a density distribution in the envelope region, which
is assumed to   correspond to the structure of an $n=1.5$ polytrope, thus  
\begin{equation}
E=3{\rho_c\over \bar \rho}{\left({5\over 2}{x_c\over 1-x_c}\right)}^{3/2},
\label{A5}
\end{equation}
where $\bar \rho = (3/  4\pi)(M_*/ R_*^3) $ is   the mean density. We obtain $\rho_c$, $x_c$, $R_*$ and $\bar \rho$
using the set of numerical stellar models described above. 

\end{appendix}

\label{lastpage}

\end{document}